\documentclass[aps,pre,twocolumn,superscriptaddress]{revtex4-1}

\usepackage[english]{babel}
\usepackage[utf8x]{inputenc}
\usepackage{amsmath,amsfonts,amssymb,amsbsy,eucal,dsfont}
\usepackage{color}
\usepackage{graphicx}
\usepackage{rotating}
\usepackage{setspace}
\usepackage{longtable}

\newcommand{\rhort}{{\rho(\vec{r},t)}}
\newcommand{\rkommat}{{(\vec{r},t)}}
\newcommand{\vecnab}{{\vec{\nabla}}}
\newcommand{\cF}{{\cal F}}
\begin{document}

\title{Dynamical density functional theory for the diffusion of injected Brownian particles}
\author{H.~L\"owen}
\affiliation{Institut f\"{u}r Theoretische Physik II,
Weiche Materie,
Heinrich-Heine-Universit\"{a}t D\"{u}sseldorf,
40225 D\"{u}sseldorf,
Germany}

\author{M. Heinen}
\affiliation{Institut f\"{u}r Theoretische Physik II,
Weiche Materie,
Heinrich-Heine-Universit\"{a}t D\"{u}sseldorf,
40225 D\"{u}sseldorf,
Germany}
\affiliation{Division of Chemistry and Chemical Engineering,
California Institute of Technology,
Pasadena, California 91125, USA.
}
\begin{abstract}
While the theory of diffusion of a single Brownian particle in confined geometries is well-established by now,
we discuss here the theoretical framework necessary to generalize the theory of diffusion to dense suspensions of strongly
interacting Brownian particles. Dynamical density functional theory (DDFT) for classical Brownian particles
represents an ideal tool for this purpose. After outlining the basic ingredients to DDFT we show that it can be
readily applied to flowing suspensions with time-dependent particle sources.
Particle interactions lead to considerable layering in the mean density profiles,
a feature that is absent in the trivial case of noninteracting, freely diffusing particles.
If the particle injection rate varies periodically in time with a suitable frequency, a resonance in the layering of the
mean particle density profile is predicted.
\end{abstract}
\maketitle
\section{Introduction}
\label{intro}
The diffusion of a Brownian particle in confined geometries such as channels,
obstacles and ratchets has been intensely studied over the last decades both by theory \cite{review_Haenggi_Marchesoni},
real-space experiment on colloids \cite{Bechinger-single_file-diffusion} and computer simulation \cite{Nielaba_EPJ_ST_2013}.
Some basic effects, such as rectification, are observed already for a single particle while others are collective such
as single-file diffusion \cite{KollmannPRL}. A large part of recent studies considers diffusion of a single Brownian particle in confining geometries. \\ \\
In this minireview, we show that dynamical density functional theory (DDFT) provides an ideal framework to study
the dynamics of interacting Brownian particles in confining geometries, and we provide a brief introduction into the DDFT equations.
The most important quantity entering into these equations is the equilibrium
two-particle direct correlation function $c(r)$ which can be determined from the pair-interaction potential $V(r)$
via liquid integral equations, based on the Ornstein-Zernike equation \cite{HansenMcDonald}. Here we outline the DDFT for diffusing profiles
of density variations on top of a homogeneous background density in an arbitrary number of spatial dimensions $d$.
Results are presented for the generic case of hard disk suspensions in $d=2$ dimensions, at various particle number densities. 
Our results include Green's functions that correspond to spatio-temporal point sources of particles,
steady-state density profiles around a constantly emitting particle source,
which correspond to a chemotactic potential \cite{Tsori_deGennesEPL,Sengupta_Kruppa=op0231},
as well as spatially and temporally oscillatory density profiles for injection of particles with
a periodically time-dependent rate.
In contrast to the smooth density profiles that are found for freely diffusing, noninteracting particles,
we find considerable particle layering for suspensions where strong particle interactions prevail.
For temporally periodic particle injection we find resonances in the density profile at a suitably chosen
particle injection frequency which matches the inverse time scale needed to advect a density peak in the flowing suspension.\\ \\ 
The remainder of this work is organized as follows:
we outline the dynamical density functional theory in Sec.~\ref{DDFT},
which is followed by a brief description of the specific systems under study in Sec.~\ref{Brown}. 
In Secs.~\ref{DDP}, \ref{Chemo}, and \ref{Periodic}, we apply DDFT to compute diffusion profiles
and chemotactic potentials for three different types of particle injection.
In Sec.~\ref{conclusion}, we sketch possible generalizations of DDFT towards other situations and draw our conclusions.
\section{Dynamical density functional theory (DDFT) \label{DDFT}}
The formulation of dynamical density functional theory \cite{TarazonaMarconi,ArcherEvans,EspanolLoewen=op0217}
(see \cite{Emmerich_Advances_inPhysics_2012=ra0033} for a recent review) starts from the continuity equation
\begin{flalign}\label{Continuity}
\frac{\partial \rho(\vec{r},t)}{\partial t} + \vec{\nabla} \cdot \vec{j}(\vec{r},t) = 0, 
\end{flalign}
for the compressible one-particle density field $\rho ( {\vec r}, t)$, which guarantees particle number conservation.
Neglecting solvent-mediated hydrodynamic interactions of the suspended particles,
the current density ${\vec j}({\vec r}, t)$ is given by Fick's law
\begin{flalign}\label{2}
\vec{j}(\vec{r},t) = -\frac{1}{\gamma} \rho(\vec{r},t) \vec{\nabla} \mu [\rho(\vec{r},t)] 
\end{flalign}
with $\mu[\rho ( {\vec r}, t)]$ denoting a formal chemical potential which is in general given as a
functional of the one-particle density $\rho ( {\vec r}, t)$, and with $\gamma$ denoting the Stokesian drag coefficient.
The chemical potential can be derived from the equilibrium free energy density functional ${\cal F}[\rho ( {\vec r}, t)]$ as
\begin{flalign}\label{3}
\mu[\rho] = \frac{\delta \cal F [\rho]}{\delta \rho}.
\end{flalign}
Hence we obtain the following DDFT equation of generalized diffusion
for the time evolution of the density field:
\begin{flalign}\label{3A}
\gamma\frac{\partial \rho(\vec{r},t)}{\partial t} = \vec{\nabla}\rho(\vec{r},t) \vec{\nabla} \left(\left.{\frac{\delta \cal F}{\delta \rho}}\right|_{\rho=\rho(\vec{r},t)}\right).
\end{flalign}
For any system with a given pairwise additive interaction, characterized by a pair-potential $V(r)$, this functional exists \cite{Mermin}
and depends parametrically on the thermal energy $k_BT$. In equilibrium, minimization of the functional
yields the equilibrium one-particle density \cite{Lowen_1994_Physics_Reports=ra0003}.
In general one can split the functional into three parts as
\begin{flalign}\label{Free_energies}
{\cal F} [\rho] = {\cal F}_{id}[\rho] + {\cal F}_{exc} [\rho] + {\cal F}_{ext}[\rho]
\end{flalign}
where
\begin{flalign}\label{Ideal_Free_Energy}
\cF_{id}[\rho] = k_BT \int d^dr~\left[\ln(\Lambda^d \rhort) - 1 \right] \rhort
\end{flalign}
is the free energy for a non-interacting (\textit{i.e.} ideal gas) system with $\Lambda = h / \sqrt{2\pi m k_B T}$
denoting the (in the following irrelevant) thermal de Broglie wavelength in terms of Planck's constant $h$, particle mass $m$, Boltzmann's constant $k_B$
and absolute temperature $T$. The exponent $d$ in Eq.~\eqref{Ideal_Free_Energy} denotes the number of spatial dimensions of the system (typically $d=1,2,3$).
The excess free energy functional ${\cal F}_{exc}[\rho ( {\vec r}, t)]$ is not known explicitly for interacting systems
and needs to be approximated. The last term in Eq.~\eqref{Free_energies} is the external free energy, which is given by
\begin{flalign}\label{6}
\cF_{ext}[\rho] = \int d^dr V_{ext}(\vec{r},t) \rhort,
\end{flalign}
where $V_{ext}({\vec r}, t)$ denotes an external potential, an
example of which is a confining, generally time-dependent geometry.\\ \\
A popular approach to construct an excess free energy functional
is the so-called Ramakrishnan-Yussouff approximation \cite{Ramakrishnan1979}
where one expands the system perturbatively around a fixed reference bulk fluid density ${\bar \rho}$ as follows:
\begin{flalign}
&\cF_{exc}[\rho] \cong -\frac{k_BT}{2} \times \nonumber\\
&\int d^d r \int d^d r' c^{(2)} ( |\vec{r}-\vec{r}' |, \bar{\rho}) (\rhort -\bar{\rho}) ( \rho(\vec{r}',t) - \bar{\rho}).\label{RY}
\end{flalign}
Here, $c^{(2)}(r, {\bar \rho}, T) \equiv c(r)$ denotes  the equilibrium direct correlation function
which is fixed by the particle interaction potential $V(r)$ at fixed uniform bulk number density ${\bar \rho}$ and temperature $T$.
One could in principle use more accurate approximations for the excess free energy functional, like fundamental measure theory
for non-overlapping particles \cite{Rosenfeld1989,Zhang2010,Roth2012,Yamani2013,Reinhardt2013,Neuhaus2013}. Nevertheless, in order to allow for
straightforward analytic progress, we choose the simpler Ramakrishnan-Yussouff scheme in the present work.   

Upon entering the approximation in Eq.~\eqref{RY}, the DDFT Eq.~\eqref{3A} becomes
\begin{flalign}\label{8}
\frac{\partial \rho(\vec{r},t)}{\partial t} =& D_0\Delta \rho(\vec{r},t) +
\vec{\nabla} \left( \rhort \frac{\vecnab V_{ext}(\vec{r},t)}{k_BT} \right) \nonumber\\ &-
\vecnab \rhort \int d^dr' \vecnab c( |\vec{r}-\vec{r}'|)(\rho(\vec{r}',t) - \bar{\rho})
\end{flalign}
with $D_0=k_BT/\gamma$ denoting the free diffusion coefficient of the
Brownian particles. At this stage, a couple of remarks are in order:
first, the DDFT equation can be derived from the Smoluchowski equation \cite{ArcherEvans}
by using one essential approximation, namely the so-called adiabatic approximation
stating that the one-particle density field is the single slow variable of the system
(i.e. all other variable are much faster) \cite{EspanolLoewen=op0217}.
Thereby nonequilibrium correlations are approximated by equilibrium correlations.
This explains why there is no additional noise term in the DDFT equations.
It is only for phase transitions that such noise terms have been phenomenologically
introduced \cite{Emmerich_Advances_inPhysics_2012=ra0033}.
Second, an alternative approach to incorporate particle interactions has been recently
proposed by Santamaria-Holek \textit{et al.} \cite{Rubi}, which works with an entropic activity coefficient and includes only local terms.\\ \\
As it stands in Eq.~\eqref{8}, the DDFT equation is nonlinear and can therefore not be solved analytically in general.
A basic insight into the physics of diffusion can, however, be gained by linearizing the DDFT equation with respect
to small density variations around the prescribed mean density ${\bar \rho}$, resulting in the following equation for the dimensionless
relative density deviation $\varepsilon ({\vec r}, t) = (\rho ({\vec r}, t) - {\bar \rho})/ {\bar \rho} $ :
\begin{flalign}\label{9}
\frac{\partial \varepsilon(\vec{r},t)}{\partial t} =
& ~D_0\Delta \varepsilon(\vec{r},t) +  \frac{1}{\gamma}\Delta V_{ext}(\vec{r},t) \nonumber\\
+&~\frac{1}{\gamma} \vecnab \left( \varepsilon(\vec{r},t) \vecnab V_{ext}(\vec{r},t) \right) \nonumber\\
-&~D_0 \bar{\rho} \int d^dr' \Delta c( |\vec{r}-\vec{r}'|)\varepsilon(\vec{r}',t).
\end{flalign}
While this equation is still hard to solve for a general external potential,
we restrict ourselves here to a time-independent external force in $x$-direction, giving rise to the external potential
\begin{flalign}\label{10}
V_{ext}(\vec{r},t) = -g x
\end{flalign}
that corresponds, for $d=2$, to a tilted plane along which the particles are allowed move (see Fig.~\ref{fig:schematic}).

Using Eq.~\eqref{10} for the external potential, Eq.~\eqref{9} simplifies to
\begin{flalign}\label{11}
\frac{\partial \varepsilon(\vec{r},t)}{\partial t} &= D_0\Delta \varepsilon(\vec{r},t) -
\frac{g}{\gamma}\frac{\partial \varepsilon(\vec{r},t)}{\partial x}\nonumber\\
&- D_0 \bar{\rho} \int d^dr' \Delta c( |\vec{r}-\vec{r}'|)\varepsilon(\vec{r}',t).
\end{flalign}
Clearly, for vanishing particle correlations  $(c(r)\equiv 0)$ the free diffusion equation is recovered. Hence, Eq.~\eqref{11}
is a generalized diffusion equation for interacting particles. Equation~\eqref{11} can be solved analytically by a Fourier transform resulting in
\begin{flalign}\label{12}
\frac{\partial \tilde{\varepsilon}(\vec{k},t)}{\partial t} =
-D_0 k^2 ( 1 - \bar{\rho} \tilde{c}(k) ) \tilde{\varepsilon}(\vec{k},t) - i \frac{g}{\gamma} k_x \tilde{\varepsilon}(\vec{k},t),
\end{flalign}
where a tilde denotes Fourier transformation in the following sense:
\begin{flalign}\label{13}
\tilde{\varepsilon}(\vec{k},t) = \int d^dr e^{-i \vec{k} \cdot \vec{r}} \varepsilon(\vec{r},t).
\end{flalign}
The first term on the right-hand-side of Eq.~\eqref{12} describes particle diffusion, and 
the second term leads to a drift induced by the external force. In the limit of diverging 
length scales (\textit{i.e.} vanishing ${\vec k}$-vectors), particle motion is governed by a collective diffusion coefficient $D_c=D_0(1-\bar \rho {\tilde c}(0))$
that is proportional to the inverse isothermal compressibility of the reference fluid \cite{HansenMcDonald}
(see also the discussion in Ref.~\cite{Hopkins_Schmidt}).
Thermodynamic stability  of the bulk reference fluid requires $1-\bar \rho{\tilde c}(0)>0$.
For repulsive interactions $V(r)$, one typically has $1-\bar \rho {\tilde c}(0)>1$, and therefore $D_c>D_0$.
\section{Brownian particles injected into a plane}
\label{Brown}
\begin{figure}
\begin{centering}
 \includegraphics[width=.95\columnwidth]{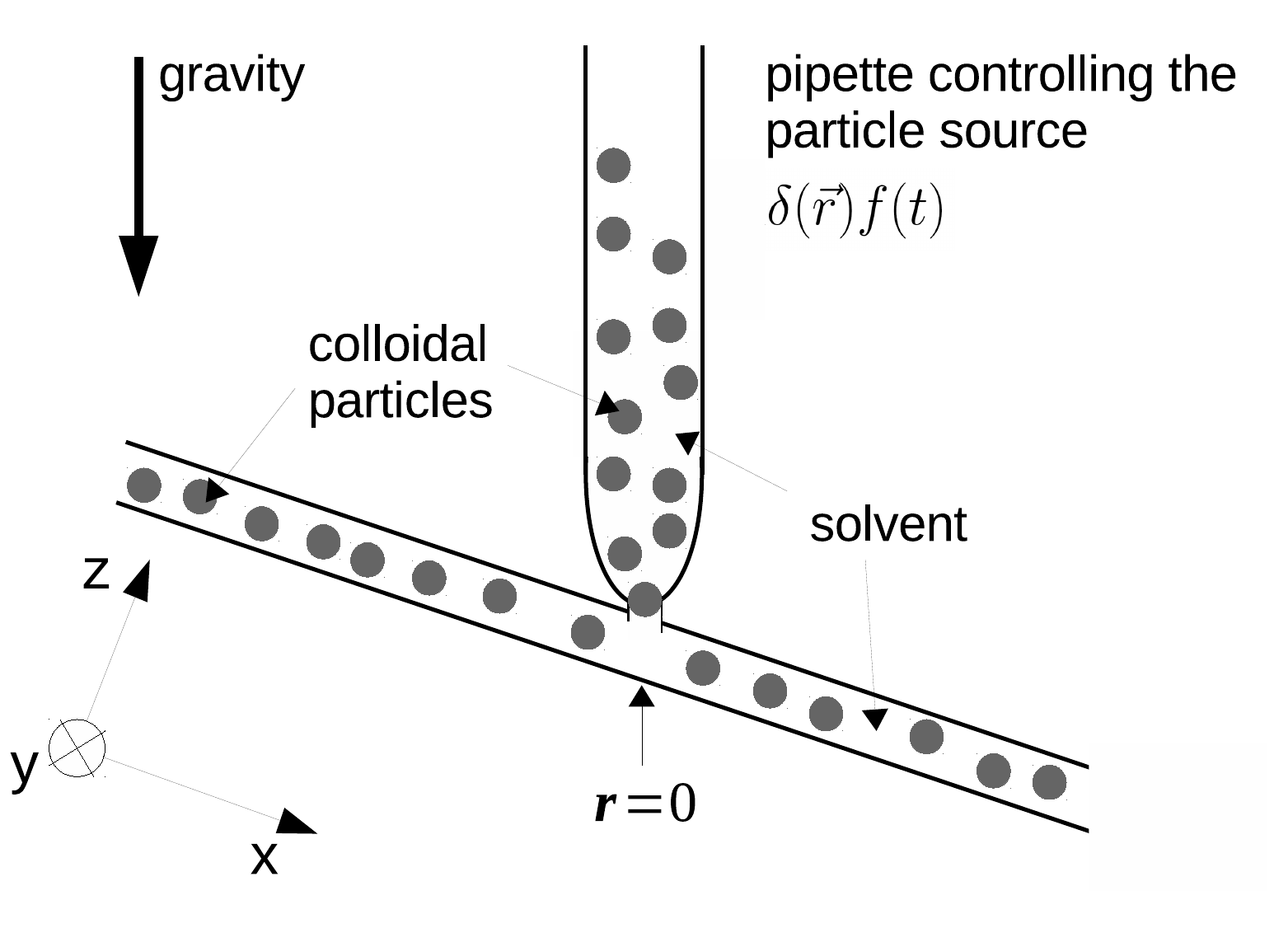}  
 \caption{Injection of colloidal particles into a tilted, flowing suspension.}
 \label{fig:schematic}
\end{centering}
\end{figure}

Figure~\ref{fig:schematic} illustrates the type of physical problem that we will consider in the following three sections.
Colloidal particles are injected at position $\boldsymbol{r}=0$ into a colloidal suspension confined to slit and driven in $x$-direction.
The nozzle diameter of the injecting pipette is only marginally 
larger than the diameter of a single colloidal particle, such that the source term that must be added to the generalized diffusion equation
is proportional to a delta function $\delta(\boldsymbol{r})$.
Different kind of time-dependencies $f(t)$ for the source term are considered in the following three sections:
In Sec.~\ref{DDP}, results are computed for a source term $f(t) = \alpha \delta(t)$, describing the instantaneous ejection of 
a number $\alpha$ of particles at time $t=0$.
A stationary, constantly emitting source term, $f(t) = \lambda = const$ is considered in Sec.~\ref{Chemo}, and
in Sec.~\ref{Periodic}, a periodic source term, $f(t) = \lambda \left[\cos(\omega t) + 1\right]$, is assumed.
The suspension injection problem studied here may be thought of as an inertia-free version
of a liquid jet impinging on a plane, a problem that has been studied in detail in experiments, and in the theoretical
framework of continuum fluid mechanics \cite{Yarin2006}.

Note here, that the formalism applied in the present work is restricted to the computation of density modulations $\varepsilon$ around
a homogeneous mean (reference) density $\bar{\rho}$. In case of a quiescent bulk fluid, or a constant external force acting on a bulk fluid
as described by the external potential in Eq.~\eqref{10}, the homogeneous mean density assumption is obviously valid. There are, however,
situations where the present formalism cannot be straightforwardly applied. One such example, where the mean fluid density is not homogeneous
but instead position-dependent, is a fluid dripping onto the apex of a paraboloid that opens downward.
Note also that the present formalism is merely a linear response field theory. The computed density deviations $\varepsilon$ represent Green's functions, linearly
weighted with the source prefactor $\alpha$ or $\lambda$. Whenever we present results for density deviations in the following,
we select source terms that cause the plotted functions to assume values typically of the order of one. Large but finite density modulations can always
be scaled down linearly (by scaling the source term) into the linear response regime. Steady state density profiles for continuous particle injection
can result in a divergence of the computed density deviation near the point of injection (see Fig.~\ref{fig:stationary}). In such cases, the present theoretical
description applies only at sufficiently large distance from the source, where the density modulation amplitudes have decreased into the linear response regime.

\section{Green's functions for diffusing density profiles}
\label{DDP}
Diffusing density profiles are obtained by putting an additional source term $\alpha \delta({\vec r})\delta (t)$
on the right hand side of Eq.~\eqref{11}. The resulting solution $\varepsilon_D({\vec r}, t)$
describes how $\alpha$ particles  initially located at the origin ${\vec r}=0$ at time $t=0$ will relax
to the final homogeneous density profile $\rho ({\vec r}, t\to\infty) = {\bar \rho}$. Experimentally this can be realized by
trapping particles with a laser tweezer \cite{EgelhaafHanes}, releasing the laser trap, and observing the
relaxation of the colloidal suspension in real-space.
Formally, the solution for $\varepsilon_D({\vec r}, t)$ is the Green's function of the generalized diffusion equation.\\ \\
By Fourier transformation one obtains
\begin{flalign}
&\varepsilon_D\rkommat =  \dfrac{\alpha}{{(2 \pi)}^d} \times\nonumber\\
&\int d^d k~e^{i \vec{k} \cdot \vec{r}}
\exp\left\lbrace{-D_0 k^2 ( 1 - \bar{\rho} \tilde{c}(k) )t - i \frac{g}{\gamma} k_x t}\right\rbrace.\label{Greensfunc}
\end{flalign}
Obviously, for vanishing correlations the standard Green's function
\begin{flalign}\label{Gaussian}
\varepsilon_D\rkommat = \dfrac{\alpha}{{(2 \pi)}^d} \left( \frac{\pi}{D_0 t}\right)^{{d}/{2}} \exp\left\lbrace{- \frac{r^2}{4 D_0 t}}\right\rbrace
\end{flalign}
for the free diffusion problem is recovered as a special case.

\begin{figure*}
\begin{centering}
 \includegraphics[width=.95\textwidth]{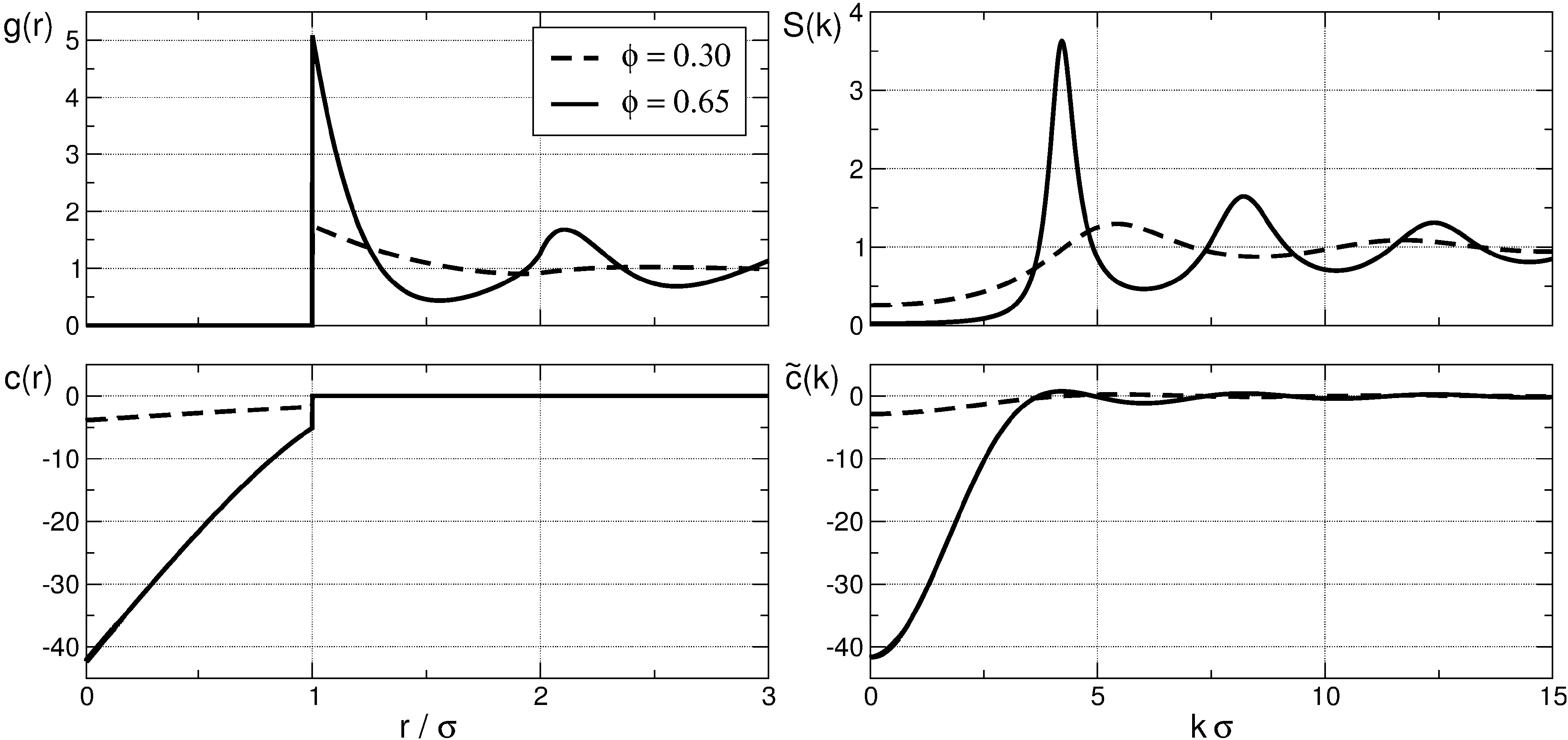}  
 \caption{Pair-correlation functions for two-dimensional homogeneous and isotropic reference fluids of hard disks at packing fractions $\phi = 0.30$ (dashed curves)
 and $\phi = 0.65$ (solid curves), computed by numerical solution of the Percus-Yevick integral equation. Panels feature the radial distribution functions
 $g(r)$ (top left), the static structure factors $S(k)$ (top right), the direct correlation functions $c(r)$ in real space (bottom left) and the direct correlation
 functions $\tilde{c}(k)$ in wavenumber space (bottom right).}
 \label{fig:PY}
\end{centering}
\end{figure*}
In order to include particle iteractions, we solve the the approximate Percus-Yevick \cite{Percus1958}
integral equation scheme
\begin{eqnarray}
g(r) - 1 &=& c(r) + \bar{\rho} \int d^d r'~c(r-r') [g(r') - 1],\label{equ:OZ}\\
c(r) &=& \left[ 1 - \exp\left\lbrace \frac{V(r)}{k_B T} \right\rbrace \right] g(r)\label{equ:PY}
\end{eqnarray}
for a homogeneous and isotropic reference fluid, to obtain a nontrivial solution for ${\tilde c}(k)$.
In Eq.~\eqref{equ:PY}, $g(r)$ is the radial distribution function \cite{HansenMcDonald},
which is not to be confused with the external drive $g$ in the present paper.

The presented DDFT equations, as well as the Percus-Yevick equation, can be numerically efficiently solved for
any physically relevant kind of isotropic, pairwise additive particle interactions, and for arbitrary spatial dimension $d$ \cite{Heinen2014}.
However, for simplicity we restrict ourselves here to the generic case of 
nonoverlapping hard disks in $d=2$ dimensions, with pair potential
\begin{equation}
 V(r) = \left\lbrace
   \begin{array}{ll}
   \infty \, & ~\mbox{for}~~~r \leq \sigma,\\~\\
   0 \, & ~\mbox{for}~~~r > \sigma,
   \end{array}
 \right. \, \label{equ:pair_pot}\\
\end{equation}
depending on the hard-disk diameter $\sigma$. 
In absence of external drive and particle sources, a two-dimensional hard disk system is governed solely by the particle packing fraction $\phi = \pi (\sigma/2)^2 \bar{\rho}$
\cite{HansenMcDonald,Book_Ivlev_Royall}.
In case of the hard disk potential, Eq.~\eqref{equ:PY} can be recast into the two conditions $c(r>\sigma) = 0$ and $g(r<\sigma) = 0$ that need to be solved
in combination with the Ornstein-Zernike Eq.~\eqref{equ:OZ}. The hard hypersphere Percus-Yevick equation has been solved (semi)-analytically
for all odd \cite{Rohrmann2007} and even \cite{Adda-Bedia2008} dimensions $d$,
but these solutions can involve considerable analytical effort and do not generally result in simple closed expressions for $\tilde{c}(k)$.
In the present work, we therefore rely on the accurate and efficient numerical solution method that we have published in Ref.~\cite{Heinen2014}.

A possible extension of the present formalism, promising higher accuracy at the price of a considerably increased numerical effort,
would consist in replacing the Percus-Yevick scheme equations \eqref{equ:OZ} and \eqref{equ:PY} for a homogeneous reference fluid by 
an inhomogeneous liquid integral equation with position-dependent number density. The resulting coupled set of DDFT and inhomogeneous
liquid integral equations can be solved self-consistently, in an approach similar to those in Refs.~\cite{Plischke1986,Kjellander1991,Nygard2012}.
In the present study, we do not apply this more accurate kind of formalism for the sake of simplicity.

In Fig.~\ref{fig:PY} we plot the (homogeneous) Percus-Yevick solutions for the pair-correlation functions of two hard-disk fluids at packing fractions $\phi = 0.3$ (dashed curves)
and $\phi = 0.65$ (solid curves). While the pair correlations are relatively weak for $\phi = 0.3$, the hard disk fluid at $\phi = 0.65$ is quite close to the
fluid-hexatic phase transition (occurring around $\phi \approx 0.70$ \cite{Engel2013}), thus exhibiting very pronounced pair-correlations. Starting with the upper left panel
and proceeding in clockwise direction, the panels in Fig.~\ref{fig:PY} feature the two rerence fluid's radial distribution functions $g(r)$, the static structure factors
$S(k) = 1 / [1 - \bar{\rho} \tilde{c}(k)]$, and the direct correlation functions in wavenumber space, $\tilde{c}(k)$, and in real space, $c(r)$. At $r = \sigma$,
both functions $c(r)$ and $g(r)$ exhibit a discontinuity at all non-zero fluid densities. 

Note that the Percus-Yevick scheme is not exact, but nevertheless very accurate in predicting the pair-correlations of hard disks in the density range studied here:
Highly accurate numerical calculations of the hard disk equation of state by Kolafa and Rottner \cite{Kolafa2006} can be used as a reference solution.
They found a normalized pressure of $p / [\bar{\rho} k_B T] = 8.39$ at packing fraction $\phi = 0.65$.
At the same packing fraction, the Percus-Yevick scheme predicts a value of $8.29$ for the normalized
hard disk virial pressure, $p_{\text{vir}} / [\bar{\rho} k_B T] = 1 + \sqrt{\pi\phi}~g(\sigma^{+})$, corresponding to an underestimation
of the reference result by $1\%$ only.

Hard disks can serve as a model system for sterically stabilized colloidal spheres that are confined to a planar surface,
and the $d=2$ systems studied here can be experimentally realized in a straightforward way, by dripping colloidal
particles onto a tilted plane (\textit{c.f.}, Fig.~\ref{fig:schematic}).
In case that the colloidal particles are not perfectly confined to a plane but 
restricted in their movement to a very narrow three-dimensional slit between parallel walls,
it has been shown \cite{Franosch2012} that lateral and transversal degrees of freedom decouple
asymptotically as the slit width vanishes, and that particle motion is governed by an effective,
slit width dependent in-plane pair potential.
\begin{figure*}
\begin{centering}
 \includegraphics[width=.8\textwidth]{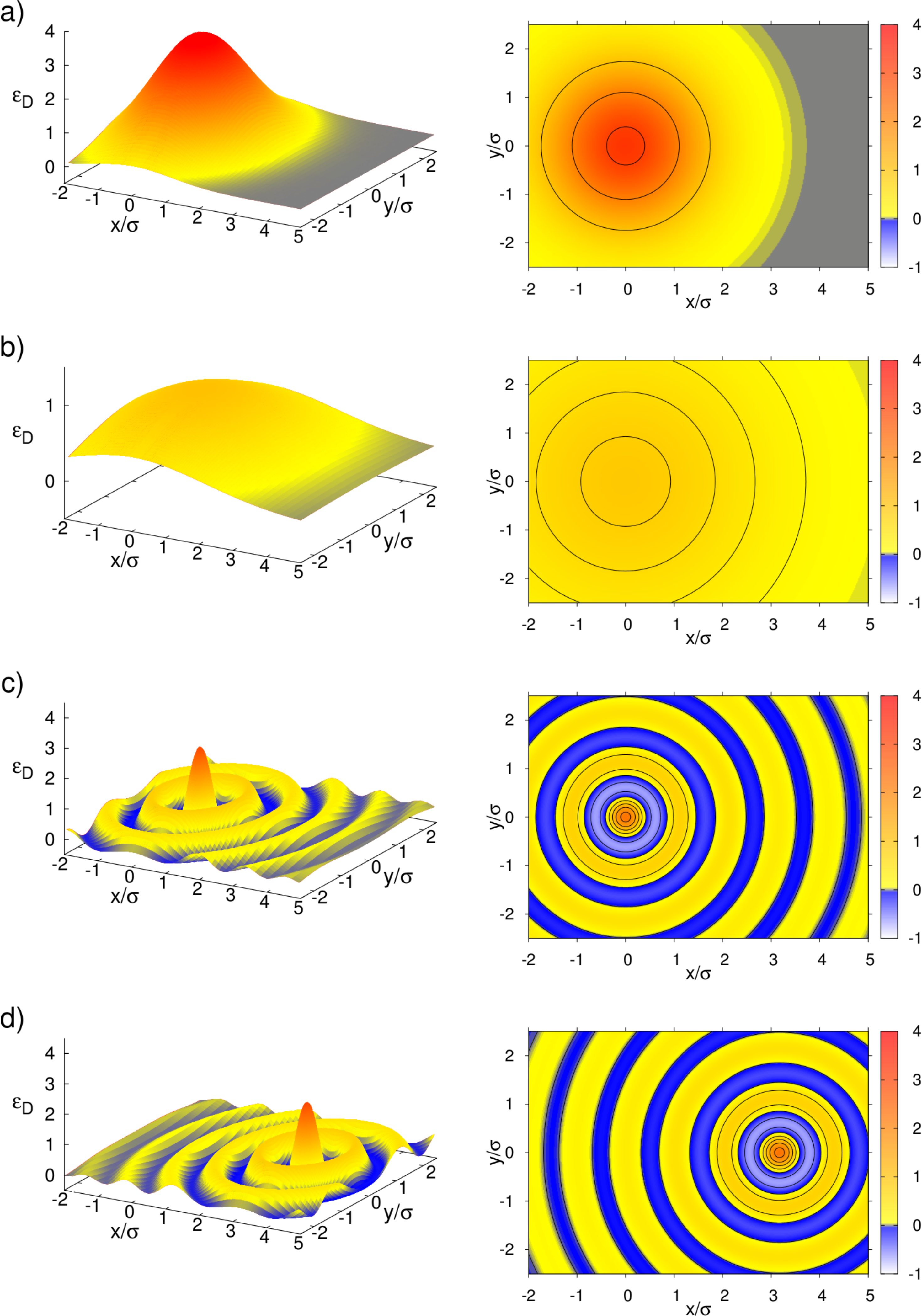}

%
%
%
%
 \caption{Green's functions $\varepsilon_D(\boldsymbol{r},t)$ for density deviations corresponding
 to a spatio-temporal particle point-source $\alpha \delta(\boldsymbol{r}) \delta(t)$.
 The top row of panels, labeled a), is for an ideal gas, and the three lower panel rows b)--d)
 are for different fluids of hard disks. Plotted are the numerical
 solutions of Eq.~\eqref{Greensfunc}.
 All results are for $d=2$ spatial dimensions, for a source strength of $\alpha = 10 / {\bar{\rho}}$,
 and all panels correspond to the same time $t = \pi {\sigma}^2 / (4.8 D_0)$.
 Ideal gas results are given by the Gaussian function in Eq.~\eqref{Gaussian}.
 Panel row b) is for a hard-disk fluid at packing fraction $\phi = 0.3$, and 
 Panel rows c) and d) are for a dense hard disk fluid at $\phi = 0.65$.
 All panels except row d) are for vanishing driving force, $g=0$.
 Results in panel row d) are for $g = 10 \times (\gamma D_0 {\bar{\rho}}^{1/2})$ and otherwise the same parameters as in row c).
 The color code is the same in all panels.
 }
 \label{fig:Greensfunc}
\end{centering}
\end{figure*}

Results for the diffusive reduced density profile $\varepsilon_D({\vec r}, t)$,
based on a numerical evaluation of Eq.~\eqref{Greensfunc}, are shown in Fig.~\ref{fig:Greensfunc}.
All input parameters for the calculations are given in the figure caption.
Each of the panels of Fig.~\ref{fig:Greensfunc} is a snapshot of the Green's function, taken always at
the same time $t = \pi {\sigma}^2 / (4.8 D_0)$ after the release of particles by the point source.
In the uppermost row a) of panels in Fig.~\ref{fig:Greensfunc}, the smooth Gaussian
profile for freely diffusing, noninteracting particles is shown.
The next lower row of panels, b), is for an interacting hard disk suspension at a moderate packing fraction of $\phi = 0.3$.
Under these conditions, the shape of the density profile remains 
quite similar to the one obtained for ideal gas free diffusion. However, the density profile in panel row b)
is spreading out quicker than the profile for free diffusion, due to the reduced osmotic compressibility of the hard disk suspension.
In addition, the density profiles for interacting particles exhibit \textit{layering}, \textit{i.e.}, regions where the radial derivative of
the density profile becomes positive. Layering is indeed present in the system plotted in panel row b) of Fig.~\ref{fig:Greensfunc}, but it is not visible
in the scale of the panels, since the the layering-induced modulations of the density profile occur too far away from the particle source,
and at too small values of $|\varepsilon_D|$. 
A significant degree of particle layering, and a further accelerated spread of the profile, is observed in row c)
of Fig.~\ref{fig:Greensfunc}, which is for packing fraction $\phi = 0.65$.
Note that there is no finite critical packing fraction for the onset of layering. Instead, due to the discontinuity in $c(r)$ at $r=\sigma$
being present at all non-zero packing fractions, the layering also occurs for all non-zero packing fractions.
Rows a)--c) of Fig.~\ref{fig:Greensfunc} all correspond to a vanishing external drive, $g = 0$, as signaled by the
location $\boldsymbol{r} = 0$ of the density profile centers.
The lowermost row d) of panels in Fig.~\ref{fig:Greensfunc} is for a non-zero driving force $g = 10 \times (\gamma D_0 {\bar{\rho}}^{1/2})$
that acts in the positive $x$-direction, and otherwise for the same parameters as in row c) of the same figure. The only effect
of the external drive is a convective displacement of the spreading density profile in $x$-direction.
\section{Chemotactic density profiles for a constant particle source}
\label{Chemo}
We now consider solutions of Eq.~(\ref{11}), with a source term $\lambda \delta({\vec r})$ added to the right hand side of the equation,
corresponding (for $\lambda > 0$) to a steady inflow of particles.
The resulting solution describes how particles flow outwards from a source with constant
ejection rate $\lambda$ located at the origin ${\vec r}=0$. 
In colloidal experiments, this can be realized by particle inflow along the third dimension
(for such an example see Ref. \cite{PalbergJCPOguz2013}).
The same kind of density profiles computed here describes the distribution of low molecular weight chemoattractants or -repellents 
in the context of chemotaxis, and therefore one might call the computed functions chemotactic profiles.
Here we denote the time-independent (steady-state) solution as $\varepsilon_C({\vec r})$. \\ \\
In the case of free diffusion in the absence of external drive ($g=0$),
the steady-state chemotactic density profile is formally identical to the solution
of the classical Poisson equation of electrostatics \cite{Tsori_deGennesEPL}.
For finite external drive, but for free diffusion only, the analytical solution has been discussed by Sengupta et al \cite{Sengupta_Kruppa=op0231}.\\ \\ 
Chemotaxis in an arbitrary concentration field $\rho({\vec r},t)$ is most easily described by
a sensing force in direction of the spatial gradient of $\rho({\vec r},t)$, such that $\rho({\vec r},t)$
can be viewed as a chemotactic potential \cite{Sengupta2}.
For an ensemble of mutually chemotactically sensing objects the solution $\varepsilon_C({\vec r})$
is therefore proportional to the chemotactic pair potential. In $d=3$,
for free diffusion and positive chemotaxis, it has been shown by Tsori and de Gennes \cite{Tsori_deGennesEPL}
that this is formally equivalent to the classical problem of gravitational collapse. \\ \\
Using Fourier transformation one obtains the expression  
\begin{flalign}\label{Stat_epsilonD}
\varepsilon_C(\vec{r}) = \dfrac{\lambda}{{(2\pi)}^d} \int d^d k ~ e^{i \vec{k} \cdot \vec{r}} \frac{1}{D_0 k^2 ( 1 - \bar{\rho} \tilde{c}(k) ) + i \frac{g}{\gamma} k_x}
\end{flalign}
for the steady state chemotactic density profile.
In the ideal gas case (that is, for $\tilde{c}(k) = 0$) the convection-diffusion equation with stationary source term can be solved analytically. The corresponding
two-dimensional ideal gas result reads $\varepsilon_C(\vec{r}) \propto \exp \left[ gx / 2\gamma D_0 \right] K_0\left[gr / 2\gamma D_0 \right]$,
in terms of $r = \sqrt{x^2 + y^2}$ and the modified Bessel function of the second kind, $K_0$.
As one can easily verify from Eq.~\eqref{Stat_epsilonD} or from the ideal gas result, the function $\varepsilon_C(\boldsymbol{r})$ diverges logarithmically to infinity when
the limit $r \to 0$ is taken. In fact, the ideal gas result diverges to infinity at every point $\vec{r}$, in the limit $g \to 0$ of vanishing external drive. This is an
artifact of assuming a two-dimensional system with one particle source, and no particle sink: At long times, the particle number density diverges everwhere.
In the Fourier integral in Eq.~\eqref{Stat_epsilonD} the divergent contribution occurs in the infinite wavelength limit $k \to 0$, corresponding to a spatially homogeneous
density that grows with no bounds. We circumvent this artifact in the following, using computational grids in $\boldsymbol{r}$- and $\boldsymbol{k}$-space that do not include
the points $\boldsymbol{r}=0$ or $\boldsymbol{k} = 0$: In our implementation of the Fast-Fourier-Transform, we employ rectangular grids with equidistant gridplanes
that are parallel to the $x$- and $y$-axis. The grid is offset from both coordinate axes by half a grid spacing, such that the divergent part of the integrand is not sampled,
leaving only the finite density modulations to be computed.

\begin{figure*}
\begin{centering}
 \includegraphics[width=.8\textwidth]{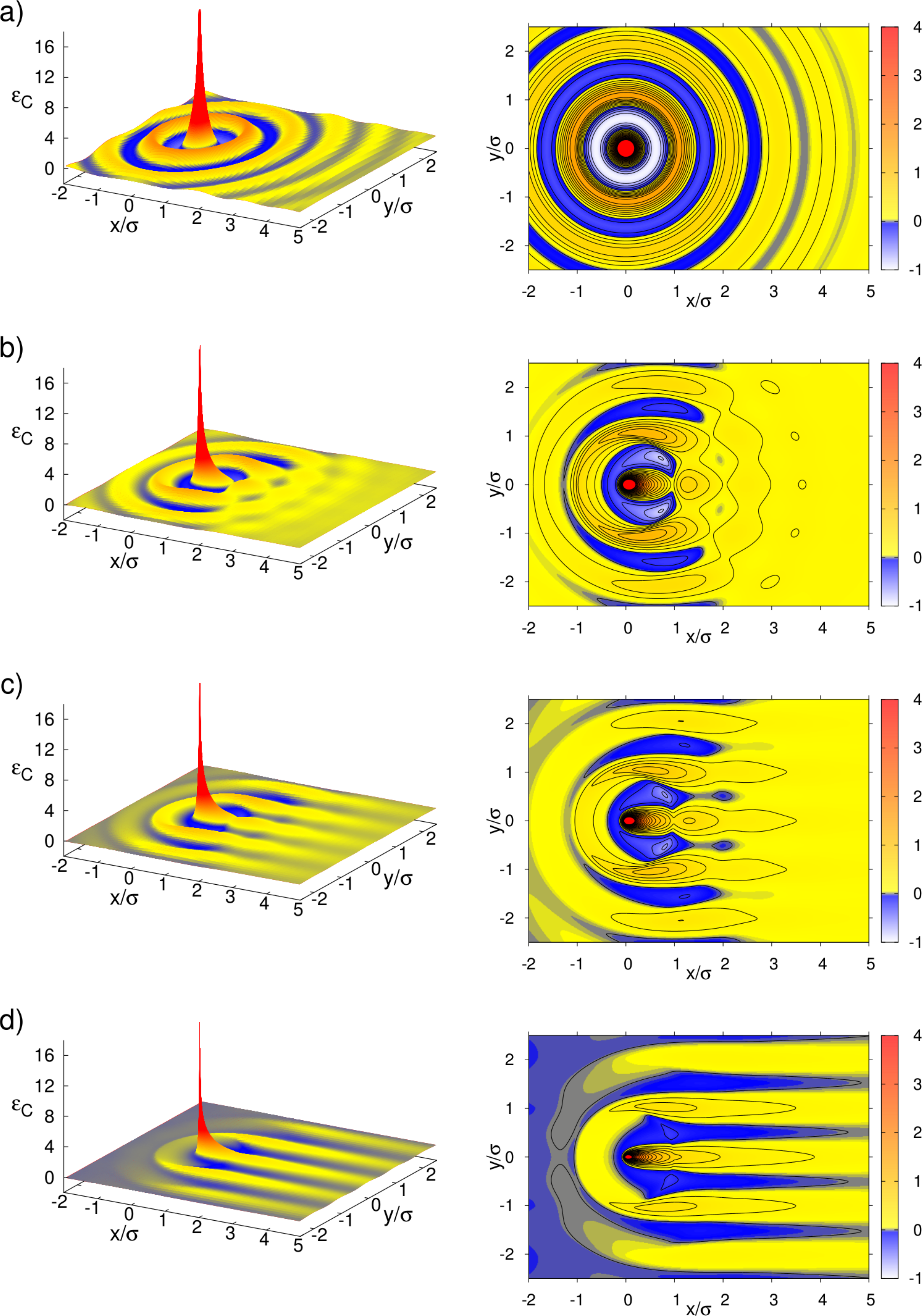} 
 
%
%
%
 
 \caption{Stationary density deviation solutions $\varepsilon_C(\boldsymbol{r})$ for
 hard disk fluids in $d=2$ spatial dimensions, at area fraction $\phi = 0.65$, obtained by numerical solution of Eq.~\eqref{Stat_epsilonD}.
 A point-like particle source, located at $\boldsymbol{r} = 0$, ejects particles at a constant, time-independent rate $\lambda = {(2\pi)}^2 D_0$.
 The results shown in the uppermost row of panels (a) are for a vanishing driving force $g = 0$.
 Results in panel rows b), c), and d) are for $g / (\gamma D_0 {\bar{\rho}}^{1/2}) = 10$, $20$, and $50$, respectively.
 The functions are sampled on grids that exclude the point $\boldsymbol{r}=0$, at which $\varepsilon_C(\boldsymbol{r})$ diverges to infinity.
 }
 \label{fig:stationary}
\end{centering}
\end{figure*}

In Fig.~\ref{fig:stationary}, explicit examples for $\varepsilon_C(\vec{r})$ are shown.
All results in Fig.~\ref{fig:stationary} are for packing fraction
$\phi = 0.65$, and for a point source of particles at $\boldsymbol{r} = 0$, ejecting particles at a constant, time-independent
rate $\lambda = {(2\pi)}^2 D_0$. The only variation in input parameters, for the results in the four rows of panels a)--d) of Fig.~\ref{fig:stationary},
is in the external drive $g$, which increases from top to bottom: In the uppermost panel row a), the external drive vanishes, and a radially
symmetric chemotactic profile is obtained around $\boldsymbol{r}=0$. For panel rows b), c), and d), we have chosen 
$g / (\gamma D_0 {\bar{\rho}}^{1/2}) = 10$, $20$, and $50$, respectively. With increasing external drive, an increasing deformation of the density
profile towards positive $x$-direction is observed, as particles get carried away from the source into that direction.
Significant layering of particles is observed in all panels of Fig.~\ref{fig:stationary}. For strong external drive
(\textit{c.f.} the lowermost row of panels), the layering results in flow-aligned high density strips that extend far away from the source
in the downstream direction. The lateral spacing between the strip centers is equal to the particle diameter $\sigma$.
For lower packing fractions (not shown in Fig.~\ref{fig:stationary}), particle layering becomes less pronounced, but 
the distance between layers remains equal to $\sigma$ at all finite packing fractions $\phi$. This can be understood on basis of Fig.~\ref{fig:PY} (lower left panel):
the discontinuity in $c(r)$ always occurs at $r = \sigma$ and varies in strength only.
In the infinite dilution limit $\phi \to 0$, the ideal gas result without layering is recovered in the numerical solution.
Note that the particle source creates an approximately semi-circular bow wave in upstream direction.
Due to the infinite speed of information transmission in the underlying classical diffusion equation, the upstream bow wave is observed at all
values of the external driving force $g$ (\textit{i.e.}, a Mach-cone is never observed).
\section{Periodic particle injection}
\label{Periodic}
Let us now consider a source term $\lambda \delta(\boldsymbol{r}) \left[\cos(\omega_0 t) + 1\right]$ added to the right-hand-side of
Eq.~(\ref{11}), corresponding to temporally oscillatory injection of particles at $\boldsymbol{r}=0$, with an ejection rate
that varies harmonically between zero and $2\lambda$, at a period of $T = 2\pi/\omega_0$.
Once again, using Fourier transformation we obtain the expression
\begin{flalign}
&\varepsilon_O(\vec{r},t) = \dfrac{\lambda}{{(2\pi)}^d} \times\nonumber\\
&\int d^d k ~ e^{i \vec{k} \cdot \vec{r}}
\left[\dfrac{1}{h(k)} + \dfrac{h(k)\cos(\omega_0 t) + \omega_0\sin(\omega_0 t)}{h^2(k) + \omega_0^2} \right]\label{Osci_epsilonO}
\end{flalign}
for the resulting, spatially and temporally oscillating density profile, where we have defined the function $h(k) = D_0 k^2 ( 1 - \bar{\rho} \tilde{c}(k) ) + i g k_x / \gamma$.

A resonance criterion for the injection angular frequency $\omega_0$ can be constructed by demanding that the product
$T v_{\text{drift}} = 2\pi g/ (\gamma \omega_0)$  should be equal to the hard disk diameter $\sigma$. Here, $v_{\text{drift}} = g/\gamma$
is the free particle drift velocity under external drive $g$. The resonance criterion can be re-cast into the form
\begin{flalign}\label{Resonance}
\omega_0 \stackrel{!}{=} \omega_0^{\text{res}} = {\pi}^{3/2} \dfrac{g}{\gamma} \sqrt{\dfrac{\bar{\rho}}{\phi}}.
\end{flalign}
Selecting an injection angular frequency $\omega_0$ that satisfies the criterion in Eq.~\eqref{Resonance} results in \textit{constructive interference}
of the injected, diffusing and convecting density profile: If Eq.~\eqref{Resonance} is satisfied, maximal ejection rate occurs at those times
at which the bow wave density maximum, created during previous peak injection, has been convectively transported downstream
to a position $\boldsymbol{r} \approx 0$, \textit{i.e.} close to the point of injection.
On the other hand, an angular frequency of $\omega_0 = 2 \omega_0^{\text{res}}$ results in \textit{destructive interference}:
In this case, peak injection rate occurs at times when the principal bow wave density \textit{minimum} arrives at $\boldsymbol{r} \approx 0$.
\begin{figure*}
\begin{centering}
 \includegraphics[width=.8\textwidth]{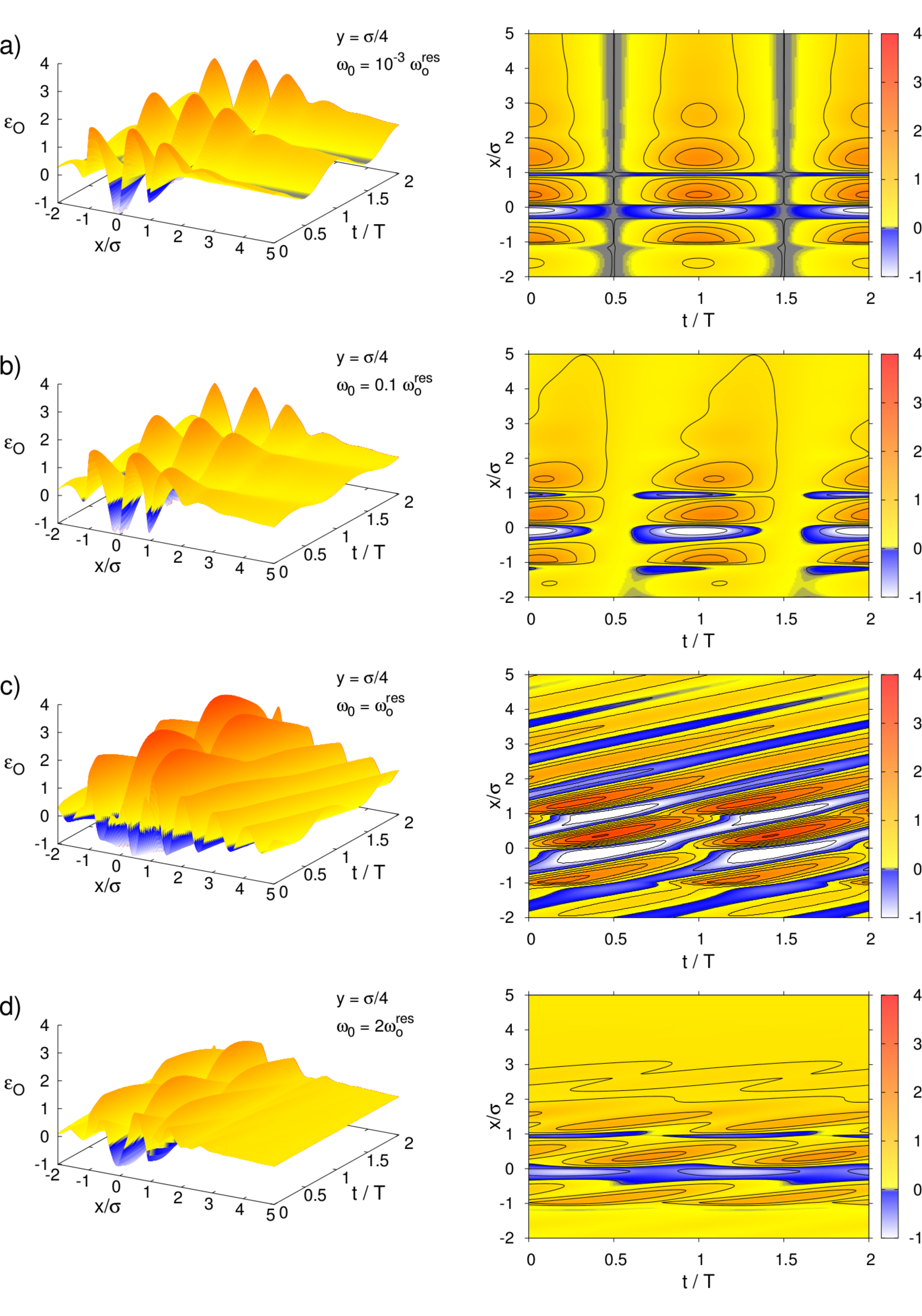}

%
%
%
 
 \caption{Time-evolution for one-dimensional slices of the oscillatory density deviation $\varepsilon_O(\boldsymbol{r},t)$,
 obtained by numerical solution of Eq.~\eqref{Osci_epsilonO}.
 A point-like particle source, located at $\boldsymbol{r} = 0$, ejects particles at a positive, time-dependent rate $\lambda \left[\cos(\omega_0 t) + 1\right]$,
 into a fluid subject to an external force $g$ in positive $x$-direction. All results shown here are for
 hard disk fluids in $d=2$ spatial dimensions, for packing fraction $\phi = 0.65$, for $\lambda = 3 \times {(2\pi)}^2 D_0$,
 and for $g = 20 \times (\gamma D_0 {\bar{\rho}}^{1/2})$.
 All plotted slices through the density function are for $y = \sigma/4$.
 Top row of panels (a): Quasi-stationary solution for a very low angular frequency of injection, $\omega_0 = 10^{-3} \times \omega_0^{\text{res}}$.
 Panel row b): Injection with a low, non-resonant angular frequency $\omega_0 = 0.1 \times \omega_0^{\text{res}}$.
 Panel row c): Resonant injection at $\omega_0 = \omega_0^{\text{res}}$, resulting in constructive interference.
 Panel row d): Anti-resonant injection at $\omega_0 = 2 \times \omega_0^{\text{res}}$, resulting in destructive interference.
 }
 \label{fig:oscillatory}
\end{centering}
\end{figure*}

In Fig.~\ref{fig:oscillatory}, we plot our numerically obtained solutions of Eq.~\eqref{Osci_epsilonO}.
Note that each panel in Fig.~\ref{fig:oscillatory} depicts the time evolution for a $y = const$ slice of the density deviation:
While one axis is for the $x$-direction, the other axis is for increasing times $t$.
As discussed in the previous section, the computational grid excludes both planes $x = 0$ and $y = 0$,
in order to avoid passing trough the divergence at $x = y = 0$. We have therefore chosen to plot slices for the small but non-zero
coordinate $y = \sigma/4$ in Fig.~\ref{fig:oscillatory}. 
All results in Fig.~\ref{fig:oscillatory} are for hard disk fluids in $d = 2$ dimensions, each of which is
subject to the same external drive, $g = 20 \times (\gamma D_0 {\bar{\rho}}^{1/2})$. Furthermore, all results
are for the same injection rate $\lambda \left[\cos(\omega_0 t) + 1\right]$ with $\lambda = 3 \times {(2\pi)}^2 D_0$, and particle
injection occurs always at $\boldsymbol{r} = 0$.
The only difference in input parameters, for the four systems depicted in Fig.~\ref{fig:oscillatory}, is in the angular
frequency of particle injection, $\omega_0$:
The top row of panels (a) in Fig.~\ref{fig:oscillatory} is for a very small angular frequency of injection, $\omega_0 = 10^{-3} \times \omega_0^{\text{res}}$.
In this case, where the injection frequency is much less than any other characteristic frequency of the suspension, a quasi-static density
profile is observed, whose $x$- and $t$-dependencies factorize almost perfectly.
In the second row of panels, b), a substantially higher angular frequency of injection, $\omega_0 = 0.1 \times \omega_0^{\text{res}}$ is chosen.
During the time period $T$, corresponding to the latter injection frequency, convective transport in positive $x$-direction can be clearly observed
to cause a tilting of the density profile in the $(x,t)$-plane (\textit{c.f.}, the right panel in row b) of Fig.~\ref{fig:oscillatory}).
However, the injection frequency in panel row b) is still very different from the resonance frequency in Eq.~\eqref{Resonance} and therefore,
the height of the peaks and the undulations for $x > 0$ remain similar to the ones observed in panel row a) of the same figure.
A qualitative difference from row a) and b) is observed in panel row c) of Fig.~\ref{fig:oscillatory}: Here, the angular frequency of particle
injection is exactly equal to the resonance angular frequency ($\omega_0 = \omega_0^{\text{res}}$), which results in excessive peaking of the density profile,
and in pronounced undulations along the downstream direction.
The lowermost row of panels, d), is for anti-resonant particle injection at an angular frequency of $\omega_0 = 2 \times \omega_0^{\text{res}}$.
As discussed above, in this case the interference of convectively transported density minima with the density function at the locus of particle
injection leads to destructive interference. The observed density profiles are less peaked that the ones in panel rows a)-c), and the undulations
die out quickly in the downstream direction.

\section{Conclusions and generalizations}
\label{conclusion}
In conclusion, we have proposed the dynamical density functional approach as a versatile
tool to address transport and diffusion of interacting, injected Brownian particles in confined geometry.
After reviewing the theory briefly we have discussed a linearized version of the equation with external drive included.
From this approximative equations some special cases of density profiles, including steady-state
profiles around a constantly emitting particle source and temporally oscillating profiles around
a source that injects particles at a periodic rate, were calculated on basis of the Percus-Yevick approximation
for the particle pair-correlations. Strong particle layering is observed, as well as resonances in the
time-dependent density profiles. Both layering and resonance are absent in the case of freely diffusing
particles that do not interact. The presented theory should be a suitable framework to address many more questions, some of which are summarized below. \\ \\
There is a plethora of possible further applications and generalizations of DDFT.
It should be noticed that DDFT can be applied to {\it{arbitrary confined geometries}} by a suitable modeling
of the external potential $V_{ext}({\vec r}, t)$ which in general, however, requires a full numerical solution of the problem.
As an example, a rather straightforward modification of the present study could consist in replacing
the tilted plane potential in Eq.~\eqref{10} by an oscillating plane potential or a
tilted ratchet potential \cite{Lichtner2010}.
Note also that {\it{hydrodynamic interactions}} between particles and particles and walls \cite{Takagi}
can be readily incorporated into the DDFT approach \cite{RexLoewenPRL2008,RexEPJE2009} on an approximate level,
at the expense of higher numerical effort for the evaluation of the transport equations.
In addition, bulk phase transitions including {\it crystallization} are in principle included in the equilibrium
functional \cite{Barrat1990,Baus1990} and therefore the dynamics of crystallization can be explored \cite{TeeffPRL,Schmiedeberg}.
Phase transitions have been intensely studied in the conceptually simpler phase-field-crystal (PFC) approach \cite{Elder,Emmerich_Advances_inPhysics_2012=ra0033,TeeffPRL},
and it would be interesting to explore the dynamics of crystallization in confined geometry, which is still an open problem.
DDFT can readily be applied to {\it mixtures}, see e.g. \cite{Lichtner2012}, and to {\it{orientational}}
degrees of freedom relevant for liquid crystals \cite{DDFT_anisotropic,Wittkowski,Brand,myLCJPCM}
where even an arbitrary particle shape has been considered \cite{shape}, including general Brownian orientational diffusion.
In addition, even more complex situations like a {\it{temperature gradient}} can be tackled by using DDFT \cite{WittDDFT}. \\ \\
Finally, {\it{active}} Brownian particles \cite{Schimansky-Geier}
can be considered and again DDFT is generalizable to this recently extensively studied nonequilibrium class of systems \cite{WensinkPRE2008,shape,Menzel}.
Experiments on  colloids \cite{Kuemmel} and bacteria \cite{Silber} have proven that the Brownian statistics with white noise
is sufficient to describe the data. Even single active particles show various new phenomena such
as circle swimming \cite{circleswimmer}, rectification \cite{Reichert}, negative mobility \cite{Cates,MarchesoniPRL}
and polar order \cite{Stark}. And we are still at the beginning to understand the fascinating collective behavior
such as turbulence \cite{PNAS}, clustering \cite{Bocquet,Buttinoni,Chaikin,BialkeEPL} and crystallization \cite{Speck,Menzel}
of active swimmers.
\section*{acknowledgement}
We acknowledge Katarina Popowa for help in typing the manuscript.
This work was supported by the ERC Advanced Grant INTERCOCOS (project
number 267499).

\end{document}